# A Survey of mmWave Backscatter: Applications, Platforms and Technologies

YIMIAO SUN, School of Software and BNRist, Tsinghua University, China
YUAN HE[∗], School of Software and BNRist, Tsinghua University, China
YANG ZOU, School of Software and BNRist, Tsinghua University, China
JIAMING GU, School of Software and BNRist, Tsinghua University, China
XIAOLEI YANG, Beijing Jiaotong University, China
JIA ZHANG, School of Software and BNRist, Tsinghua University, China
ZIHENG MAO, School of Software and BNRist, Tsinghua University, China

As a key enabling technology of the Internet of Things (IoT) and 5G communication networks, millimeter wave (mmWave) backscatter has undergone noteworthy advancements and brought significant improvement to prevailing sensing and communication systems. Past few years have witnessed growing efforts in innovating mmWave backscatter transmitters (*e.g.,* tags and metasurfaces) and the corresponding techniques, which provide efficient information embedding and fine-grained signal manipulation for mmWave backscatter technologies. These efforts have greatly enabled a variety of appealing applications, such as long-range localization, roadside-to-vehicle communication, coverage optimization and large-scale identification. In this paper, we carry out a comprehensive survey to systematically summarize the works related to the topic of mmWave backscatter. Firstly, we introduce the scope of this survey and provide a taxonomy to distinguish two categories of mmWave backscatter research based on the operating principle of the backscatter transmitter: modulation-based and relay-based. Furthermore, existing works in each category are grouped and introduced in detail, with their common applications, platforms and technologies, respectively. Finally, we elaborate on potential directions and discuss related surveys in this area.

CCS Concepts: • **Hardware** → **Wireless integrated network sensors**; • **Networks** → **Sensor networks**.

Additional Key Words and Phrases: Millimeter Wave, Backscatter, Wireless Sensing, Internet of Things, RF Computing

---
[∗]Yuan He is the corresponding author.

---

Authors' addresses: Yimiao Sun, School of Software and BNRist, Tsinghua University, Beijing, China, sym21@mails.tsinghua.edu.cn; Yuan He, School of Software and BNRist, Tsinghua University, Beijing, China, heyuan@mail.tsinghua.edu.cn; Yang Zou, School of Software and BNRist, Tsinghua University, Beijing, China, zouy23@mails.tsinghua.edu.cn; Jiaming Gu, School of Software and BNRist, Tsinghua University, Beijing, China, gjm20@mails.tsinghua.edu.cn; Xiaolei Yang, Beijing Jiaotong University, Beijing, China, 21211026@bjtu.edu.cn; Jia Zhang, School of Software and BNRist, Tsinghua University, Beijing, China, j-zhang19@mails.tsinghua.edu.cn; Ziheng Mao, School of Software and BNRist, Tsinghua University, Beijing, China, mzh23@mails.tsinghua.edu.cn.

---

## 1 INTRODUCTION

Past decades have witnessed the rapid advancement of Internet-of-Things (IoT) and 5G networks [7, 13, 44, 64, 88, 132, 133], whereas the high reliance on power supplies significantly limits the ubiquitous deployment of IoT devices. Attributed to the battery-free property, backscatter has emerged as a promising technology to address this issue [34, 49, 66, 68, 91, 126, 156]. Different from traditional communication architecture, a battery-free device called the backscatter transmitter[1] (*e.g.,* backscatter tags and metasurfaces[2]) is introduced. It can harvest energy from the incident signals and then modulate or relay signals. In this way, the need for batteries is eliminated, significantly boosting the potential of low-power and ubiquitous communication and sensing.

The early efforts on backscatter technology require a dedicated transmitter to generate the carrier signal, such as the widely used Radio-frequency identification (RFID) [69, 70, 114, 133] as well as Wi-Fi-based [30, 101] and LoRa-based [65, 66] backscatter. Then, ambient backscatter attracts more interest because the signals from existing systems can act as excitation signals (*e.g.,* TV [107] and FM [132]), greatly reducing deployment and maintenance costs. However, due to the constrained channel bandwidth, existing backscatter techniques usually are significantly limited in two aspects: low data rate and low spatial resolution. On the one hand, the data rate of most existing backscatter techniques is less than 1 Mbps [140], only sufficient to support a part of low-end IoT sensors for data exchange. However, with the rapid demand for applications requiring high data rates, such as virtual reality (VR) and augmented reality (AR), the performance of these backscatter techniques remains inadequate. On the other hand, their spatial resolution (*e.g.,* range resolution) is also limited, typically 7.5 m for Wi-Fi [123, 134, 147] and even worse for RFID and LoRa. Thus, they are hard to support fine-grained backscatter-based sensing tasks unless with more complicated system designs, which in turn severely restricts their availability.

The main reason for the above-mentioned two limitations is the constrained channel bandwidth. To address these limitations, exploring millimeter wave (mmWave)-based backscatter is a promising solution. mmWave technology [11, 19, 27, 47, 111, 149, 150] has been rapidly developed and attracted extensive attention in recent years. The mmWave spectrum, ranging from 24 GHz to 300 GHz, offers an abundance of available bandwidth, making it an attractive candidate for satisfying the ever-growing demand for higher data rates and fine-grained sensing applications [26, 31, 76, 86, 96, 121, 151]. Besides, due to the shorter wavelength of mmWave signal, the corresponding backscatter transmitters can be fabricated in a miniaturized form [51], which can be easily embedded in size-restricted devices and infrastructures. Thereby, increasing efforts have focused on the topic of mmWave backscatter to explore the improvement in the performance of existing backscatter communication and sensing. Existing research on mmWave backscatter has achieved over 1 Gbps [99] data rate, thousands of times more than that of sub-6 GHz. Besides, mmWave backscatter has also been applied in autonomous driving [25] and intelligent inventory [81] for fine-grained sensing.

Moreover, the exploration of mmWave metasurfaces to fine-grained manipulate the mmWave signals has greatly released the potential of mmWave backscatter technologies. By precisely controlling the response of mmWave signals at a subwavelength scale, metasurfaces not only broaden the scope of mmWave backscatter technologies, but also enable novel and appealing applications in mmWave communication and sensing, *e.g.,* coverage expansion [1, 111] and on-vehicle sensing [104, 138].

---

[1]The term "backscatter transmitter" is a widely used designation [24, 128] in backscatter technologies. In this paper, we use "backscatter transmitters" to collectively refer to both backscatter tags and metasurfaces.
[2]Tags are typically simple and compact devices with a few antennas (*e.g.,* Van Atta Array, Luneburg Lens, Leaky Wave Antenna, *etc.*). Metasurfaces are artificially engineered surfaces composed of arrays of subwavelength elements.

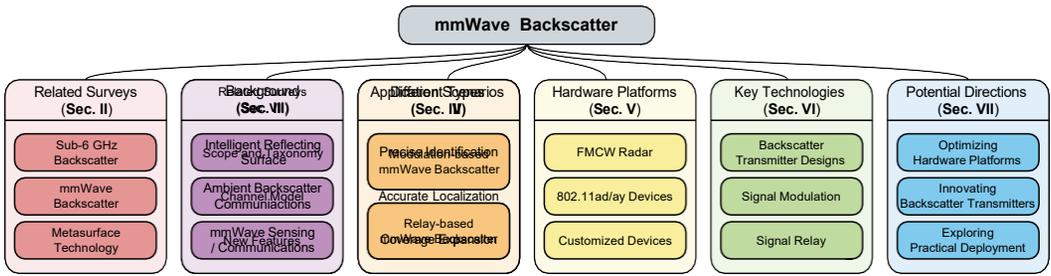

Fig. 1. Outline and roadmap of this survey on mmWave backscatter.

Despite these advantages, mmWave backscatter faces unique technical challenges in the design of backscatter transmitters and extraction of backscatter signals, due to the high-frequency feature of mmWave signals. Even though existing mmWave backscatter works have attempted to overcome these challenges to enable novel paradigms of communication and sensing in IoT, these works still keep quite a gap from the ubiquitous and practical deployment, in terms of the system performance, the design difficulty, the deployment cost, *etc*. Thus, it is important to figure out the challenges they face and the limitations they have not addressed. Moreover, the emerging applications and key technologies should be pointed out to illuminate the development trends of mmWave backscatter. Therefore, there is a strong need for a comprehensive study of mmWave backscatter.

To this end, we conduct a systematic survey of the mmWave backscatter technology, including its existing applications, platforms, and technologies as well as the challenges and potential directions. One of the cores of mmWave backscatter is the backscatter transmitters, whose operating mode largely determines the backscatter system's architecture, techniques and application scenarios. As the backscatter transmitters in existing works either modulate the incident mmWave carrier signal or only act as a relay of the incident signal, we categorize the existing works into two types based on these two modes, that is modulation-based mode and relay-based mode.

The main contributions of this paper are summarized as follows:

- We present a complete picture of the literature in the area of mmWave backscatter and also propose a novel taxonomy that originates from the operating mode of backscatter transmitters which guides the first step.
- The application scenarios, hardware platforms and key technologies of mmWave backscatter are summarized, respectively. Wherein, we elaborate on the key technologies of each category according to our novel taxonomy.
- Following a detailed discussion of the literature, we comprehensively discuss the key challenges and future directions related to mmWave backscatter, including optimizing hardware platforms, innovating backscatter transmitters and exploring practical deployment.

As shown in Fig. 1, the structure of this survey is organized as follows. Sec. 2 summarizes related surveys and points out the salient novelties of our paper. Then, Sec. 3 introduces the background and taxonomy of this survey. After which, Sec. 4 briefly introduces different types of mmWave backscatter based on our taxonomy. Sec. 5 discusses mmWave hardware platforms, which always act as the transmitter (Tx) or the receiver (Rx) in backscatter architectures. Sec. 6 elaborates on the key technologies of mmWave backscatter from backscatter transmitter design to signal modulation and relay. Based on the analysis and summary of these works, Sec. 7 discusses potential challenges and future directions. We conclude this survey in Sec. 8.

Table 1. Summary of related surveys

| Referenc | Backscatter Transmitter | Carrier Signal | Taxonomy |
|---|---|---|---|
| Van *et al.* [128] | Tag | sub-6 GHz (TV, FM, Cellular, Wi-Fi, *etc.*) | Operation modes of ambient RF sources |
| Jiang *et al.* [67] | Tag | sub-6 GHz (TV, FM, Wi-Fi, LTE, LoRa, *etc.*) | Fundamental issues related to backscatter communication |
| Niu *et al.* [103] | Tag | sub-6 GHz (Wi-Fi, TV, FM, LoRa, *etc.*) | Form of signal sources |
| Liu *et al.* | | sub-6 GHz (TV, RFID, Cellular, Wi-Fi, BLE, *etc.*) | Architectures of next-generation backscatter system |
| Wu *et al.* [140] | Tag | sub-6 GHz (TV, FM, Wi-Fi, LTE, LoRa, *etc.*) | Form of signal sources |
| Galappaththige *et al.* [35] | Tag | sub-6 GHz (LoRa, BLE, Wi-Fi, RFID, *etc.*) | Different types of backscatter communication |
| Rezaei *et al.* [115] | Tag | sub-6 GHz (LoRa, FM, Wi-Fi, RFID, *etc.*) | Different coding schemes in backscatter communication |
| Toro *et al.* | | sub-6 GHz (RFID, LoRa, Wi-Fi, FM, *etc.*) and Visible Light | Performance enhancing techniques in backscatter-based sensing |
| Chen *et al.* | Tag [24] | | Beam alignment techniques in mmWave backscatter systems |
| Magbool *et al.* [95] | Metasurface | sub-6 GHz and mmWave | ISAC systems assisted by metasurfaces |
| Gong *et al.* [39] | Metasurface | sub-6 GHz, mmWave and THz | Roles of metasurfaces in metasurface-assisted wireless networks |
| Ahmed *et al.* [5] | Metasurface | sub-6 GHz, mmWave and THz | Different modes of active metasurface |
| **This Survey** | **Tag and Metasurface** | **mmWav** | Operating modes of mmWave backscatter transmitters |

## 2 RELATED SURVEYS

There have been several related surveys focusing on particular scopes of backscatter technologies. We summarize these surveys in Table 1 and briefly introduce these related surveys in this section. Then, we elaborate on the differences between this survey in comparison to these surveys. Finally, we point out the distinctive contributions of this survey.

Jiang *et al.* [67] focus on the integration of backscatter communication and battery-free IoT. They discuss the backscatter prototypes based on the fundamental issues of backscatter communication, including performance enhancement, concurrent transmission, security guarantee and system interplay. Liu *et al.* [87] provide a review on the next-generation backscatter communication. Beyond the conventional backscatter technology (*e.g.,* RFID), they introduce novel backscatter technologies based on their architectures, including multiple-access backscatter, ad hoc backscatter, ambient backscatter and cross-technology backscatter. Niu *et al.* [103] conduct a survey on backscatter communication from the perspective of the ubiquitousness of the carrier signal and give a detailed discussion on the potential techniques in the future backscatter communication systems. Galappaththige *et al.* [35] provides a survey and analysis on the link budget of backscatter communication, and discusses the key parameters of the communication channel, passive tags and readers. Rezaei *et al.* [115] reviews the coding techniques for backscatter communication and identifies the potential coding schemes and multiple access schemes. Both Van *et al.* [128] and Wu *et al.* [140] survey the research works on the topic of ambient backscatter communication, which is one of the widely-studied sub-area of backscatter communication. Besides the topic of communication, Toro *et al.* [127] review backscatter communication-based wireless sensing and elaborate on techniques for sensing performance enhancement, *e.g.,* power management, signal quality and channel modeling. However, these surveys mainly focus on the backscatter technologies in sub-6 GHz, paying little attention to the unique features and challenges of backscatter technologies in the mmWave band.

Some other surveys focus on metasurface technologies. Magbool *et al.* [95] pay attention to the topic of metasurface-assisted Integrated Sensing and Communications (ISAC), with an emphasis on the two levels of integration: radio-communications co-existence (RCC) and dual-function radar-communications (DFCC). Gong *et al.* [39] introduce applications of metasurfaces in wireless communications, then overview different performance metrics and analytical approaches to characterize the performance improvement of metasurface-assisted wireless networks. Ahmed *et al.* [5] provides a review of active Reconfigurable Intelligent Surface (ARIS), particularly emphasizing current improvements and its various uses within the context of 6G wireless networks. However, these surveys mainly focus on the metasurface without incorporating the concept of backscatter technology. Thus, they ignore the techniques and challenges associated with other types of backscatter transmitters, *i.e.,* various backscatter tags.

The work closest to ours is the survey conducted by Chen *et al.* [24], which summarizes the various mmWave backscatter communication techniques. They categorize these works based on the approaches to solve the beam alignment problem, *i.e.,* retrodirectivity-based and others. They introduce the principles of mmWave backscatter and the corresponding systems. Then, related applications as well as open issues and challenges are discussed. Compared to that work, our survey is more systematic and comprehensive in these four features: 1) We present a more structured and clear taxonomy to categorize existing mmWave backscatter and ensure the completeness of the taxonomy. 2) We provide a more systematic summary of the related research works including metasurface-based approaches and the latest works, which are not covered in Chen *et al.*'s work [24]. 3) We introduce more detailed theoretical and technical content to offer readers a more complete picture and deeper understanding of mmWave backscatter technologies, including the channel model, hardware platforms, and backscatter transmitter designs, *etc.* 4) We conduct an in-depth discussion on the existing challenges and potential directions of mmWave backscatter technologies based on our survey and insights.

To summarize, our survey presents novelty in the following aspects:

- Our survey is more inclusive and contains the latest advances in the field of mmWave backscatter technologies.
- Our survey presents a novel taxonomy of mmWave backscatter research based on the operating principles of backscatter transmitters.
- We conduct a comprehensive summary of the existing works and provide a detailed introduction to the key technologies in mmWave backscatter, including mmWave transmitter design, signal modulation, and signal relay.
- Our survey discusses the key challenges and future directions, which may inspire researchers and developers to conduct further research in mmWave backscatter to realize low-power, high-performance and ubiquitous communication and sensing in IoT and 5G networks.

## 3 BACKGROUND

In this section, we introduce the background of this survey. We first clarify the scope of this survey and our taxonomy to categorize the existing works. Then, we briefly introduce the basic channel model of the mmWave backscatter. Finally, we discuss the difference between mmWave backscatter and traditional sub-6 GHz backscatter.

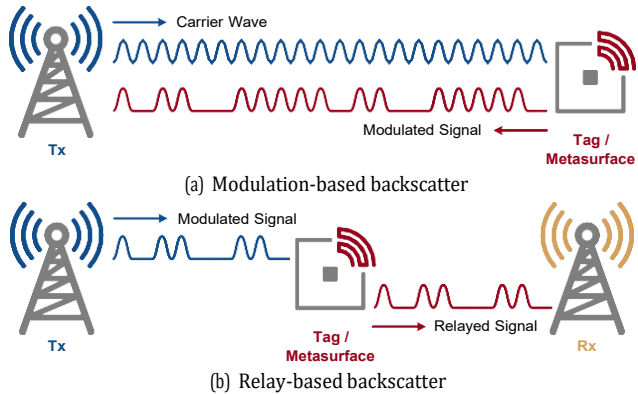

(a) Modulation-based backscatter

(b) Relay-based backscatter

Fig. 2. Different operating principles of backscatter transmitters.

### 3.1 Scope and Taxonomy

The scope of this survey contains the application scenarios, hardware platforms, and key technologies related to mmWave backscatter. mmWave backscatter is the communication technology in which the mmWave signal generated by the Tx can be modulated and reflected by a backscatter transmitter (*i.e.,* either a tag or metasurface), back to the Tx or forward to the Rx. In this process, the energy source is usually the Tx, while the information source can be either the Tx or the backscatter transmitter, indicating that the source of energy and information can be separated from each other. This characteristic is very different from that of traditional communication systems. Thus, there exist different architectures of mmWave backscatter and operating principles of backscatter transmitters: modulating the incoming signal then retracing it back to the Tx or only working as a relay to directly forward the signal to the Rx.

Based on the above discussion, existing research works on the topic of mmWave backscatter can be divided into two categories according to the operating principles of backscatter transmitters: **modulation-based** and **relay-based**.

. **Modulation-based** mmWave backscatter is a communication architecture in which a backscatter transmitter modulates the information onto the incoming signal and reflects it back to the Tx. The illustration of this architecture is shown in Fig. 2(a).
. **Relay-based** mmWave backscatter is a communication architecture in which a backscatter transmitter directly reflects the incoming signals from the Tx to the Rx without modulation[3]. The illustration of this architecture is shown in Fig. 2(b).

In the following, we will discuss application scenarios, hardware platforms and key technologies related to existing works, respectively, to inspire the exploration of new types of mmWave backscatter architectures.

### 3.2 Channel Model

Similar to the conventional communication systems, the simplest channel model of mmWave backscatter can be formulated as follows

---
[3]This process is called "backscatter-aided relay communication" in some of the exiting works [36, 40–42], indicating that the signal is backscattered and relayed by the backscatter transmitter without modulation. In this paper, we use the term "relay-based backscatter" to refer to this process.

$$s_{rx}(t) = H * s_{tx}(t) \tag{1}$$

where $s_{rx}(t)$ is the received signal by Rx and $s_{tx}(t)$ is the transmitted signal by Tx. $H$ denotes the channel response between Tx and Rx.

However, the most significant difference in the channel model between the conventional communication systems and mmWave backscatter is the double path loss and the impact of reflectivity introduced by the backscatter transmitters. By taking these impacting factors into consideration, Eq. (1) can be rewritten as

$$s_{rx}(t) = H_2(t) * \{\Gamma \cdot [H_1(t) * s_{tx}(t)]\}, \tag{2}$$

where $H_1(t)$ is the channel response between Tx and the backscatter transmitter, and $H_2(t)$ is the channel response between the backscatter transmitter and Rx. $\Gamma = |\Gamma| e^{\angle \Gamma}$ denotes the reflection coefficient of the backscatter transmitter, and is determined by the design of the backscatter transmitter and the modulation scheme.

As Eq. (2) indicates, on one hand, the backscattered signal will be impacted by the reflectivity of the backscatter transmitters. For instance, backscatter transmitters with on-off keying (OOK) modulation can alternate the magnitude of incident signals by changing $|\Gamma|$. Besides, the units of metasurfaces or phased arrays with phase-shifting ability can manipulate the phase of incident signals by changing $\angle \Gamma$. On the other hand, the signal will experience a double path loss, which is characterized by $H_1(t)$ and $H_2(t)$ as follows

$$H_1(t) = \sqrt{G_{tx}G}\frac{\lambda}{d_1}h_1(t), \quad H_2(t) = \sqrt{G_{rx}G}\frac{\lambda}{d_2}h_2(t), \tag{3}$$

where $\lambda$ is the wavelength of the signal. $G_{tx}$, $G_{rx}$ and $G$ are the gains of the Tx antennas, the Rx antennas and the backscatter transmitter, respectively. $d_1$ and $d_2$ represent the distances from the Tx to the backscatter transmitter, and from the backscatter transmitter to the Rx, respectively. $h_1(t)$ and $h_2(t)$ are the small-scale fading components.

Here, we consider the line-of-sight (LoS) path and non-line-of-sight (NLoS) paths (*i.e.,* multipath reflections) to model $h_1(t)$ and $h_2(t)$. Specifically, a Rician model [3, 63] can be deployed, so the channel response of $h_1(t)$ and $h_2(t)$ are represented as

$$h_i(t) = \sqrt{\frac{K_i}{K_i+1}} e^{-j\phi_{i,\text{LoS}}} \delta(t - \tau_{i,\text{LoS}}) + \sqrt{\frac{1}{K_i+1}} \sum_{l=1} \alpha_{i,l} e^{-j\phi_{i,l}} \delta(t - \tau_{i,l}), i \in \{1, 2\}, \tag{4}$$

where $K_i$ is the Rician $K$-factor. $\phi_{i,\text{LoS}}$ and $\tau_{i,\text{LoS}}$ are the phase and delay of the LoS signal, respectively. $\alpha_{i,l}$, $\phi_{i,l}$ and $\tau_{i,l}$ are the amplitude attenuation, phase and delay of the $l$-th NLoS signal, respectively.

Thus, the overall mmWave backscatter channel then becomes:

$$H(t) = H_2(t) * [\Gamma \cdot H_1(t)]. \tag{5}$$

Putting Eq. (2) and Eq. (3) together, the end-to-end relationship between the power of the transmitted and received signal can be modeled as:

$$P_{rx} = P_{tx} G_{tx} G_{rx} G^2 |\Gamma|^2 \frac{\lambda^4}{d_1^2 d_2^2} |h_2(t) * h_1(t)|^2, \tag{6}$$

where the terms $G^2 |\Gamma|^2$ and $1/d_1^2 d_2^2$ also indicate the unique impact of reflectivity and double path loss in mmWave backscatter systems.

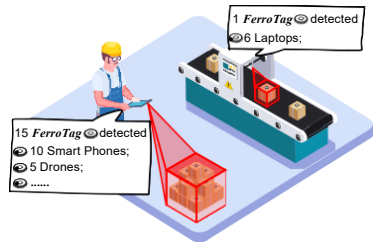

Fig. 3. Application scenes of FerroTag.

### 3.3 New Features of mmWave Backscatter

Different from the widely researched sub-6 GHz backscatter, the emerging mmWave backscatter can enable novel applications, but poses unique challenges. The following briefly introduces the new features of mmWave backscatter, compared with the traditional sub-6 GHz backscatter technology.

- **Operating mode**: The operating mode of mmWave backscatter is different from that of sub-6 GHz backscatter. On one hand, the high path loss of mmWave signals makes them very sensitive to NLoS scenarios. Thus, relay-based mmWave backscatter is very necessary in this case, while such relay-based functionality is generally not imperative in low-frequency sub-6 GHz scenarios. On the other hand, such high path loss leads to a scarcity of ambient mmWave signals. Thus, different from sub-6 GHz backscatter, it is very difficult to achieve ambient backscatter in the mmWave band.
- **Application scenarios**: Attributed to the large bandwidth of mmWave signals, mmWave backscatter can enable applications demanding high-speed transmission (*e.g.,* VR/AR) and high-resolution sensing (*e.g.,* identification and localization). However, the operation range of mmWave backscatter may be limited, while sub-6 GHz backscatter technologies have more advantages in providing long-range and broad-coverage services.
- **Main technical problems**: mmWave backscatter presents unique technical challenges compared to existing sub-6 GHz backscatter. On one hand, mmWave signal is subject to significant attenuation when propagating through both the air and backscatter transmitters, leading to constraints on the energy of the backscatter signal and potential implications for the robustness and performance of mmWave backscatter systems. On the other hand, as the signal frequency increases to the mmWave band, the design of backscatter transmitters becomes notably more intricate than that of backscatter transmitters in lower frequencies. Many components for signal processing are difficult to design and integrate into mmWave backscatter transmitters.

## 4 TYPES OF MMWAVE BACKSCATTER

mmWave backscatter is a key enabling technology for many communication and sensing applications. We can categorize them into either modulation-based or relay-based according to the architecture of their transceivers and backscatter transmitters. For example, precise localization of the backscatter transmitter always requires its retro-reflectivity to reflect the signal back to Tx, with some modulation for more precise results. It is scarcely possible to relay the query signal to another receiver for calculating the location. However, to cover the blind spots without signal coverage only requires the backscatter transmitter to resteer the mmWave beam, because the incident signal has already been modulated by Tx (*e.g.,* base station).

Following we will briefly introduce the two different types of mmWave backscatter and their typical applications.

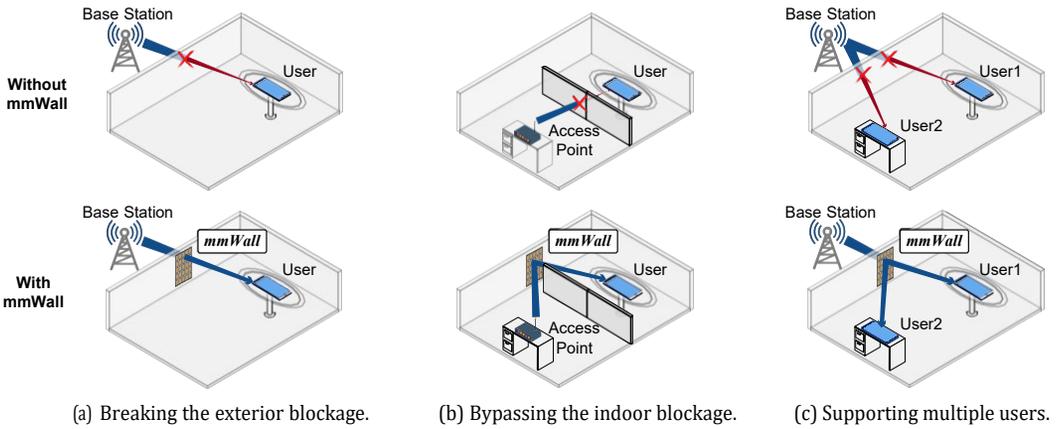

(a) Breaking the exterior blockage.  (b) Bypassing the indoor blockage.  (c) Supporting multiple users.

Fig. 4. Functionality of mmWall.

### 4.1 Modulation-based mmWave Backscatter

*4.1.1 Precise Identification.* Attributed to the high frequency, mmWave signal is sensitive to the variation in the reflection object, and thus can convey much information about the object. In this case, some works try to embed information in backscatter transmitters with unique structures, and then extract the information from the backscatter signals [81, 104] for identification.

**FerroTag** [81] attaches different tags on boxes of cargo to replace RFID for inventory, which is shown in Fig. 3. These tags are printed by ferrofluidic ink and have unique responses to mmWave signal as the printed pattern changes. This mmWave backscatter-based identification scheme can ensure an identification accuracy of more than 99 % among 201 different tags. **RoS** [104] employs passive antenna array-based tags with different stacks and numbers to represent road signs, ensuring that autonomous vehicles can identify them even in bad weather. RoS embeds information in the Radar Cross Section (RCS) which can be decoded at the radar in the vehicle.

*4.1.2 Accurate Localization.* There is plentiful bandwidth in mmWave frequency so high accuracy and low latency can be provided for localization applications. Combined with retro-reflective tags, the high path loss of mmWave signal can be compensated, thus a higher Signal-to-Noise Ratio (SNR) can be obtained. As such, the state-of-the-art method, **Hawkeye** [11], has achieved 7 mm median localization accuracy at 160 m range. Moreover, by assigning different tags with unique signatures, large-scale tags can be localized concurrently (100 in practice and 1024 in theory). **SuperSight** [10] presents the first NLoS localization scheme for mmWave backscatter. By exploiting multiple circularly polarized retro-reflective backscatter tags, the 3D localization and orientation of the target can be estimated with only less than 1 cm and $0.3°$ error, respectively. Both of the aforementioned approaches focus on localizing mmWave backscatter tags, yet each presents distinct advantages and limitations, leading to different applicable scenarios. Specifically, Hawkeye excels in large-scale, long-range tag localization with high scalability and accuracy. However, it may fail when LoS paths between the radar and tags are blocked. Consequently, Hawkeye is suitable for large-scale and rapid localization in open outdoor environments, such as urban infrastructure management and agricultural equipment monitoring. In contrast, SuperSight operates effectively in indoor NLoS scenarios with obstructions and reflections, yet may encounter challenges with mobile reflectors or without metal reflectors to provide reflection paths. Therefore, SuperSight is suitable for NLoS localization in complex indoor environments containing metal reflectors, such as industrial inventory monitoring and smart home management.

10	Y. Sun, et al.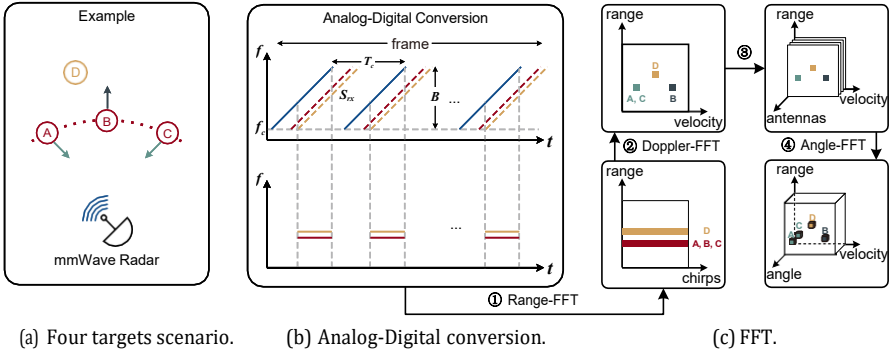

(a) Four targets scenario.  (b) Analog-Digital conversion.  (c) FFT.

Fig. 5. The sensing scenario and signal process using FMCW radar.

### 4.2 Relay-based mmWave Backscatter

*4.2.1 Coverage Expansion.* Despite mmWave signal has lots of advantages, it is fragile to blockage, resulting in spotty coverage to be a fundamental problem. Further, traditional outdoor-to-indoor coverage is almost impossible with mmWave in this case, particularly in the urban canyon. Therefore, researchers try to develop backscatter transmitters to relay mmWave signal for coverage expansion.

For example, **mmWall** [28] develops a metasurface to tackle the above-mentioned challenges. The functionality of mmWall is depicted in Fig. 4. Firstly, mmWall can refract incident signals from outdoors to steer them directly towards an indoor Rx, breaking the blockage of exterior building walls [129] and making outdoor-to-indoor communication possible. Secondly, mmWall can reflect the beam of the incident signal at non-specular angles [143, 152] to bypass the indoor blockage. And thirdly, mmWall can backscatter the outdoor signals at non-specular angles, and then support multiple indoor users at the same time.

## 5 MMWAVE HARDWARE PLATFORMS

As crucial roles in mmWave backscatter, Tx and Rx are often composed of various mmWave hardware platforms, such as Frequency Modulated Continuous Wave (FMCW) radar, 802.11ad/ay devices or even customized devices, to meet requirements of different scenarios. In this section, we briefly introduce the common mmWave hardware platforms and compare different products.

### 5.1 FMCW Radar

The FMCW radar is one of the most commonly used radars, the signal chirp of which is modulated in frequency, linear changing over a defined time, as Fig. 5(b) illustrates. Specifically, each chirp is characterized by a duration of $T_c$, a start frequency of $f_c$, and a bandwidth of $B$. Thanks to the large bandwidth and subtle phase changes, FMCW radar is widely used in many applications.

To receive and analyze the FMCW signal, the received signal $S_{rx}$ will be mixed with the original transmitted signal $S_{tx}$ at first, to form the intermediate frequency (IF) signal, whose frequency is equal to the difference between the frequency of $S_{tx}$ and $S_{rx}$. Then, the IF signal will be sampled by Analog-to-Digital Converter (ADC) and delivered to digital signal processing for further processing.

According to different communication and sensing tasks, one can selectively execute pre-processing of fast Fourier Transform (FFT) to analyze the received signal in the spectrum domain, including Range-FFT, Doppler-FFT and Angle-FFT [116]. Considering a sensing scenario shown in Fig. 5(a): There are four targets, where target D is static and far away from the radar while targets A, B and C are at the same distance to the radar as well as the same velocity. However, target B is moving away while targets A and C are with the same radial velocity towards the radar.

Table 2. Comparison of FMCW platforms

| Product | Frequency (GHz) | Tx × Rx | Sampling rate (msps) | Tx power (dBm) | Typ. Power Consumption (W) | Cost (USD) |
|---|---|---|---|---|---|---|
| IWR1443 [59] | 76-81 | 3 × 4 | 37.5 | 12 | 1.73-2.1 | ~299 |
| IWR6843 [60] | 60-64 | 3 × 4 | 25 | 12 | 1.19-1.75 | ~175 |
| AWR1243 [57] | 76-81 | 3 × 4 | 37.5 | 12 | 1.62-2.01 | ~299 |
| AWR1642 [58] | 76-81 | 2 × 4 | 12.5 | 12 | 1.3-2.14 | ~299 |
| Eval-Tinyrad [55] | 24-24.25 | 2 × 4 | 1.2 | 8 | 3.9 | ~2214 |
| Eval-DEMORAD [56] | 24-24.25 | 2 × 4 | 1.2 | 8 | ~1.79 | N/A (obsolete) |
| Radarbook2 [61] | 24-24.25 | 2 × 8 | 40 | N/A | N/A | N/A |
| Distance2Go [4] | 24-26 | 1 × 1 | 10 | 11 | 2 | ~204 |

FMCW radar can distinguish these four targets according to their distance, velocity and direction, respectively. Fig. 5 shows the detailed pre-processing flow.

. Firstly, the IF signal of each chirp is sampled and retained as a 1-D array. After receiving consecutive chirps, several 1-D arrays are combined to form a 2-D matrix, as Fig. 5(b) depicts.
. Then, Range-FFT can be executed by performing FFT along the time axis of the matrix to obtain the range spectrum (step ①). The range spectrum will exhibit two strips, in which target A, B and C are represented by the first strip, while the second strip corresponds to target D, according to their distance to the radar. Thus, target D can be identified.
. After that, FFT will be applied again along the chirp number axis, referred to as Doppler-FFT [122], yields a range-Doppler spectrum (step ②). By doing so, the first strip in the range spectrum is split into 2 blocks. The right one corresponds to target B, characterized by its positive velocity relative to the radar, as it moves away. Targets A and C, however, share the left block, since they are moving toward the radar with the same speed.
. Finally, Angle-FFT [113] will be performed to distinguish target A and C. Generally, FMCW radars comprise multiple receiving antennas, thereby forming an antenna array, so that the Angle-of-Arrival (AoA) information of each target can be obtained. This is achieved by arranging the Range-Doppler spectrum of each antenna in a 3-D coordinate system (step ③) and then applying FFT along the Rx number axis (step ④), and that is how Angle-FFT works. Until now, four targets can be separated totally.

Based on the above analysis, researchers can selectively execute the pre-processing according to their task of sensing or communication. For instance, **Millimetro** [122] applies Doppler-FFT to the backscatter FMCW signals to distinguish tags via their unique modulation frequencies, and then estimates tags' direction by applying Angle-FFT. Similarly, **R-difucial** [31] exploits Range-FFT and Doppler-FFT to distinguish the modulation patterns of different tags. Besides, FMCW signals can also help signal detection and decoding at tags. **MilBack** [89] sends "V-shape" FMCW signals so that the tag can detect the delay of signals with the same frequency for signal detection and frequency identification. **BiScatter** [106] sends FMCW signals with different slopes to encode different bits, and then the tag uses two delay lines of different lengths to decode the FMCW signals. Table 2 summarizes and compares common commercial off-the-shelf (COTS) FMCW radars to provide readers with an intuitive understanding of their abilities and differences.

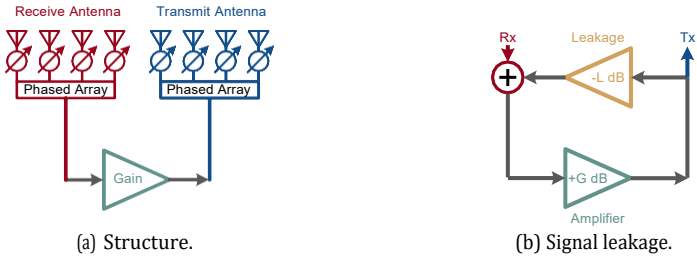

(a) Structure.  (b) Signal leakage.

Fig. 6. Illustration of the phased array–based tag.

## 5.2 802.11ad/ay Devices

60 GHz mmWave technology has become a feasible method to achieve high data rate on Wi-Fi systems. However, as signal propagation at the 60 GHz band significantly differs from that at the sub-6 GHz bands, 802.11ad redefines the fundamental principles of Wi-Fi Systems and takes advantage of high bandwidth and high directivity of 60 GHz to realize a Gbps-level transmission rate [102]. Subsequently, as emerging applications such as AR/VR and automotive driving require higher throughput, the IEEE 802.11 Task Group introduces 802.11ay, such as Multiple Input Multiple Output (MIMO), 128/256 Quadrature Amplitude Modulation (QAM) and Orthogonal Frequency-Division Multiplexing (OFDM) to achieve a 100Gbps-level transmission rate [37]. Besides, these technologies can also gain great potential for mmWave sensing applications.

There are a few mmWave backscatter works implemented based on 802.11ad/ay devices in the current literature. Most of them [18, 19, 111] utilize the Airfide 802.11ad 60 GHz radio [6], which has 8 phased array panels arranged in a $2 \times 4$ layout. Each panel consists of $6 \times 6$ quasi-omni antenna elements, with 6 dBi gain and 2-bit phase shifter per element. The 4 corner elements are disabled, so only 32 elements are usable in effect. The phase shifters can be reconfigured to form up to 128 beam patterns, each corresponding to a weight vector in a predefined codebook [135].

Currently, the 802.11ad/ay devices usually work as base stations and try to communicate with user ends, but may face the NLoS scenarios where mmWave signals are easily blocked. Thus, many existing works aim to relay these signals for adequate mmWave communication. For instance, **MilliMirror** [111] exploits passive metasurface to relay the signals from mmWave base stations to indoor users. Besides, **mmComb** [19] proposes to embed tag's information in beamforming training frames that are frequently exchanged between 802.11ad/ay APs and clients to help accommodate mmWave backscatter tags into existing mmWave WiFi networks.

## 5.3 Customized Devices

In addition to COTS mmWave hardware platforms, many researchers also conduct experiments by developing customized devices. **UniScatter** [112] carries out experiments on 60 GHz by using the ISAC device [50] from Qualcomm. This ISAC device reuses an existing 802.11ad/ay communication system as a radar sensor and provides channel impulse respondance (CIR) measurements to decode the backscatter signal. **FerroTag** [81] prototypes their Tx and Rx based on an ordinal 24 GHz FMCW mmWave system with a bandwidth of 450 MHz, which is equipped with two $4 \times 4$ Tx array for transmitting and receiving, respectively. **MiRa** [2] builds a fullfledged mmWave software-defined radio platform with electronically steerable antenna arrays on the 24 GHz ISM band. The design of MiRa enables all USRP-GNU functions to be performed in mmWave band.

## 6 KEY TECHNOLOGIES

In this section, we elaborate on the key technologies related to mmWave backscatter. We first introduce the basic but crucial component in the architecture of mmWave backscatter transmitters, by focusing on some of the most commonly used designs. Then, modulation-based and relay-based works are detailedly discussed based on their corresponding key technologies, signal modulation and signal relay. Signal modulation technologies are further divided into four categories, including frequency shift keying, on-off keying, spatial coding and others. Similarly, signal relay technologies are divided into two categories, including reconfigurable patterns and constant patterns.

Table 3. Summary of the Typical mmWave Backscatter Transmitters

| Backscatter Transmitter | Beam Pattern | Energy Cost | Typical Fabrication | Advantage | Typical Application Scenario |
|---|---|---|---|---|---|
| Phased Array | Dynamic | High | PCB | High adaptivity | Beam steering and beam scanning |
| Van Atta Array | Fixed | Low | PCB | Passive structure Low | Retro-reflection |
| Luneburg Lens | Fixed | Low | 3D printing | Passive structure Low cost 3D FoV | 3D or multi-beam retro-reflection |
| Leaky Wave Antenna | Fixed | Low | PCB | Passive structure Low | MIMO Frequency-agile scnearios |
| Metasurface | Dynamic/Fixed | Moderate | PCB 3D printing | Diverse signal property control | Beam scanning Polarization Switching Signal property control |

## 6.1 Backsactter Transmitter Designs

What sets backscatter communication apart from traditional communication is the various backscatter transmitters. They can either work as an information source to modulate the incident signal or as a relay to steer the modulated signal. Different transmitter designs own distinct abilities to manipulate signals, so proper designs could be chosen to deal with corresponding challenges in mmWave communication and sensing. Several most used backscatter transmitters in existing works will be introduced in the following and we summarize them in Table 3.

*6.1.1 Phased Array.* At mmWave frequencies, traditional antenna designs face challenges due to increased free space path loss and atmospheric absorption. Thus, phased arrays are proposed to overcome these limitations [136, 141, 153]. Phased arrays comprise multiple antenna elements that are electronically controlled to collectively shape and steer the incident signal. As shown in Fig. 6(a), **MoVR** [1] proposes a typical phased array-based tag, where both transmit and receive antennas are composed with phased arrays, and then connected via a variable-gain amplifier. Unlike traditional fixed antennas, this tag (*i.e.,* phased array) can dynamically adjust the phase and amplitude of each antenna element's signal in a few microseconds, to control both the angles of incident and backscatter signals, allowing for precise beamforming. In that case, it is promising to establish and maintain a reliable communication link even in challenging environments. Besides, this design of tag neither decodes the signal nor includes any transmit or receive digital components, thus, avoiding complex and expensive components operating at multiple Gbps.

However, a significant challenge in designing such a tag stems from the leakage between the transmit and receive arrays, which is shown in Fig. 6(b). The reason is that some backscatter signals will be received again by the receive array, thereby generating garbage signals. To deal with this problem, the amplification gain is often controlled to limit the signal leakage.

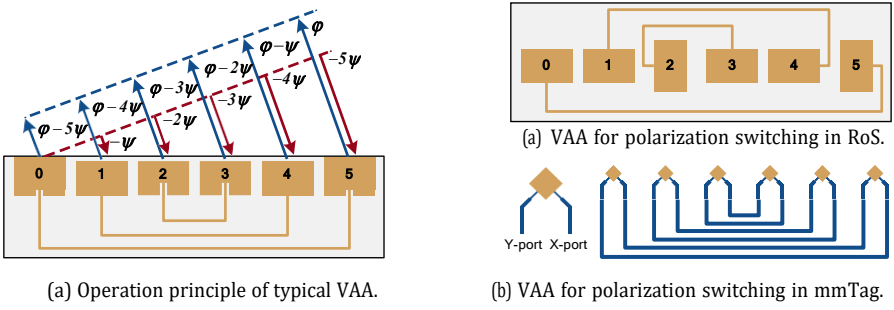

(a) Operation principle of typical VAA.

(a) VAA for polarization switching in RoS.

(b) VAA for polarization switching in mmTag.

Fig. 7. Typical structures and functions of Van Atta Array.

*6.1.2 Van Atta Array.* As we mentioned above, mmWave transceivers have to use directional antennas to focus their transmitted and received power into narrow beams to compensate for the pass loss. Besides, communication between two mmWave devices is only possible when their beams are well-aligned. This is more difficult when mmWave devices are mobile. In that case, the energy cost of phased arrays may be unaffordable for the low-power tag [46, 98]. Thus, some of the existing works tackle this problem by exploiting the Van Atta Array (VAA) [99, 104, 122].

VAA was invented in the 1960's, and named after its inventor, physicist John Van Atta [120]. As shown in Fig. 7(a), a basic VAA consists of a linear array of antenna elements with equal spacing of $\frac{\lambda}{2}$. The symmetric elements are interconnected by transmission lines (TLs). Signals received by each antenna are propagated through the TLs and re-radiated by its connected peer on the other end. Suppose an incident far field wavefront induces a phase offset, $\psi$, between adjacent antennas. The incident signal phase at the $k$-th antenna, relative to that at the 0-th antenna, is $-k\psi$. By setting the lengths of each TL to differ by multiples of $\lambda_g$ (*i.e.,* the wavelength of the signal guided in TLs), a constant wrapped phase offset $\phi$ is introduced for all signals propagating through the TLs. Consequently, given the number of antennas $N$, the $k$-th antenna receives and re-rediates the signal from the $(N-1-k)$-th antenna, whose phase relative to the 0-th antenna is $k\psi$, which is reversed compared to that of the arrived incident signal. As such, VAA can achieve retro-reflection.

However, as the backscatter signal is retro-reflected, it is hard for Tx to separate it from the incident signal. Both **mmTag** [99] and **RoS** [104] exploit polarization switching in the design of tags to cancel this interference. Specifically, mmTag decomposes the incident signal into two orthogonal directions (shown in Fig. 7(c)), and then rotates the propagation direction of these two signals by 90°, respectively. After signal superposition to form the backscatter signal, the polarization direction of the backscatter signal will be orthogonal to that of the incident signal, so as to avoid signal self-interference. Similarly, RoS rotates half of the linear polarized antennas in the VAA by 90° (shown in Fig. 7(b)) to achieve this goal.

*6.1.3 Luneburg Lens.* Traditional antenna arrays are typically composed of patch antennas or microstrip antennas, which usually support limited bandwidth due to the size and principle of metallic conductors. However, commercial mmWave radar occupies a wide frequency band, ranging from 24 GHz to 81 GHz, so traditional antenna arrays may suffer from poor availability and scalability across different applications or mmWave platforms. Moreover, most traditional antenna arrays are linearly arranged, resulting in that they can not provide wide angular field-of-view (FoV) in both the azimuth and elevation directions. Fortunately, the Luneburg Lens (LL) [90] can well address both two problems, working in a wide frequency range and FoV.

LL is a spherically symmetric gradient-index lens with permittivity profile $\epsilon(r) = 2 - \frac{r}{R}^2$, where $R$ is the radius of the lens, and $r$ is the distance of any point to the sphere center. The permittivity distribution of LL is shown in Fig. 8(a). In practical fabrication, these gradient-index lenses are

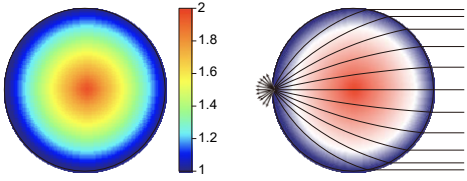
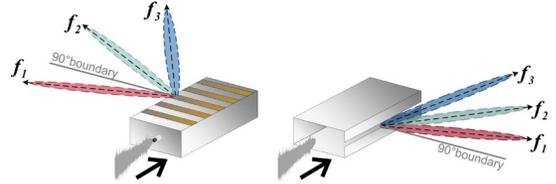

(a) Permittivity distribution. (b) Signal propagation in LL.

Fig. 8. Operation principle of the Luneburg lens.

(a) Periodic LWA  (b) Uniform LWA

Fig. 9. Radiation patterns of different LWAs.

normally layered structures of discrete concentric shells, each with a different refractive index. Signals parallelly incident to the lens will focus on a point on the sphere like Fig. 8(b) shows. The advantage of LL is two-fold. On the one hand, it can retro-reflect the incident signal in both the azimuth and elevation directions and concentrate the signal energy on a specific direction. On the other hand, its (Radio Frequency) RF response is consistent at all frequencies due to the frequency-independent permittivity profile. Thus, it has been exploited for many applications, such as designing beamforming antennas [74, 80], cooperating with antenna arrays [97, 117] and so on.

However, as LL is shaped in a sphere, a LL-based tag may be not convenient to be deployed or cascaded with other RF devices (*e.g.,* modulators). In order to align LL and modulator, **UniScatter** [112] reshapes the traditional structure to have a flat-bottom focal plane. Therefore, the proposed LL can work well with the modulator surface while retaining its retro-reflectivity.

*6.1.4 Leaky Wave Antenna.* As the demand for MIMO is growing [43, 48, 154], multiple narrowband beams are expected rather than a wideband backscattered signal. Leaky Wave Antenna (LWA) naturally meets this expectation so it is used in many mmWave backscatter applications.

LWA belongs to the class of traveling-wave antennas, where the propagating wave inside the antenna structure can "leak" (*i.e.,* radiate) from the waveguide to free space, hence the name. It can distinctively couple the frequency and radiation direction of the leaky wave (*i.e.,* the backscatter signal) to exhibit the feature of frequency and spatial division multiplexing (FSDM), as shown in Fig. 9. Specifically, the direction of the backscatter signal with frequency $f$ can be determined by [62]: $\theta(f) = \arccos \frac{\beta(f)}{k_0(f)}$, where $\beta(f)$ and $k_0(f)$ are the phase constant along the LWA and the propagation constant in free space [100]. Traditional LWA usually employs a metallic waveguide with a slit cut, namely uniform LWA, which is depicted in Fig. 9(a). The backscatter signal of uniform LWAs only propagates towards the forward region (*i.e.,* [0°, 90°]). Uniform LWAs benefit from the simple fabrication process and are mostly used in THz applications [38, 71, 72].

**GigSky** [76] adopts another type of LWA, periodic LWA, as the tag to achieve mmWave MIMO. This type of LWA is typically designed using a dielectric substrate with a periodic array of metal strips (*i.e.,* slots) [118, 142, 157] and similar to an antenna array, as shown in Fig. 9(b). The backscatter signal of this periodic LWA can propagate towards both forward and backward regions (*i.e.,* [0°, 180°]) [62]. Periodic LWA has been widely studied in recent research due to its versatile slot design and low-cost fabrication using the printed circuit board (PCB) process.

*6.1.5 Metasurface.* In some scenarios, fine-grained manipulation of mmWave beams is a critical requirement where beams are expected to be steered towards other directions rather than retro-reflected [53]. In this case, backscatter transmitters that are linearly arranged or can only retro-reflect the incident signal are no longer adequate. Metasurface [12, 23, 83], an innovative structure, has been proposed and extensively researched to meet the above-mentioned requirement.

As shown in Fig. 10, metasurface is typically a planar array of the subwavelength-sized structure [32, 78, 79] that is meticulously designed to manipulate the properties of the incident signal.

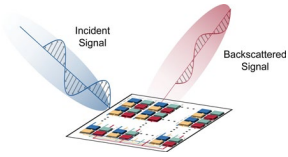

Fig. 10. Metasurface that manipulates incident signals.

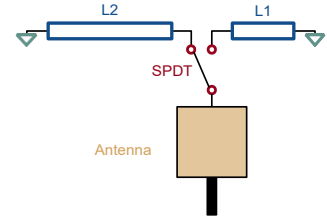

Fig. 11. SPDT in the tag of mmTag.

Table 4. Summary of the Works of Modulation-based mmWave Backscatter

| Works | Modulation Scheme | Backscatter Transmitter Design | Application Scenario |
|---|---|---|---|
| **Millimetro** [122] | OOK | Van Atta Array | Localization and identification |
| **Platypus** [52] | OOK | Van Atta Array | Micro-displacement sensing |
| **mmTag** [99] | OOK | Van Atta Array | Communication |
| **MilBack** [89] | OOK | Leaky Wave Antenna | Communication and localization |
| **R-fiducial** [31] | OOK | Van Atta Array | Localization and identification |
| **REITS** [82] | OOK | Van Atta Array | Communication |
| **BiScatter** [106] | OOK/FSK | Van Atta Array | Communication and localization |
| **Hawkeye** [11] | FSK | Van Atta Array | Localization and identification |
| **OmniScatter** [9] | FSK | Van Atta Array | Communication |
| **UniScatter** [112] | FSK | Luneburg Lens | Communication |
| **Osprey** [109] | Spatial Coding | Aluminum strips | Tire wear measurement |
| **RoS** [104] | Spatial Coding | Van Atta Array | Identification |
| **FerroTag** [81] | Spatial Coding | Metamaterial | Identification |
| **MilliSign** [54] | Spatial Coding | Corner Reflector | Identification |
| **mmComb** [19] | Others | High-Gain Antenna | Communication |
| **MetaWave** [25] | Others | Metamaterial | Attacking |
| **ThermoWave** [20] | Others | Metamaterial | Temperature monitoring |

The fundamental principle underlying metasurface is the interaction between signals and the subwavelength structures arranged in precise geometries. By tailoring the size, shape, and arrangement of these subwavelength elements, metasurfaces can control the phase [83], amplitude [8], and polarization [23] of the incident or backscatter signals.

Metasurface can be designed in either active or passive mode. **Tan** *et al.* [125] proposes an active metasurface, which is composed of a micro-controller, RF switches and antennas. This metasurface can dynamically control the direction of backscatter mmWave signals, but may be bulky to carry or deploy. Besides, existing active metasurfaces require PCB fabrication, which may be very expensive for mmWave due to the specialized substrate and strict fabrication tolerances. In contrast, **MilliMirror** [111] proposes a passive metasurface, which is fabricated using commodity 3D printers, significantly lowering the cost by at least an order of magnitude. However, the disadvantage of this type of metasurface is that once they are fabricated, the radiation pattern can not be changed.

## 6.2 Signal Modulation

One of the key functions of backscatter transmitters is working as the source of information and modulating the incident carrier wave. This is usually the core design of most modulation-based mmWave backscatter works. Table 4 summarizes these works, among which various modulation schemes are exploited to satisfy the different requirements of corresponding application scenarios, such as on-off keying (OOK), frequency-shifting keying (FSK), spatial coding and so on. The following will introduce these modulation techniques and corresponding works respectively.

*6.2.1 OOK.* OOK is a basic modulation scheme in backscatter technology [18, 82, 144, 155, 159] to enable energy-efficient and low-cost communication. At the core of OOK involves the simple binary modulation of a carrier signal. It is achieved by selectively absorbing or reflecting the incident signal to convey information. When the tag absorbs the incident signal, it is in the "off" state, representing one binary value ("0" or "1"), while reflecting the signal corresponds to the other.

Traditional backscatter systems implement OOK modulation by the RF switch on the tag. Specifically, the RF switch is used to connect the antenna to the ground plane. When the switch is on, the antenna is connected to the ground and does not reflect the incident signal, and when the switch is off, the antenna reflects the incident signal. Unfortunately, this simple approach sometimes does not work well because RF switches have high leakage at mmWave frequencies when the switch is off, and hence the the signal reflection will not change much.

To solve this problem, **mmTag** [99] use Single Pole Double Throw (SPDT) switches, where the input port of each SPDT is connected to an antenna, and the two output ports of SPDTs are connected to two transmission lines with different length, as shown in Fig. 11. The length of these lines is carefully chosen such that when the antenna is connected to line L1, the antenna is tuned, and hence it resonates and reflects the signal back. On the other hand, when the antenna is connected to the line L2, the antenna is not tuned and therefore does not reflect the signal back. In this case, by turning the switches on or off, the tag can change the amplitude of the reflected signal and the Rx can simply decode the node's data by examining the amplitude of each symbol.

**Millimetro** [122] finds that when tags modulate the incident signal in the OOK scheme, the square wave across chirps (*i.e.,* the backscatter signal) will be converted to a sinc function after range-Doppler FFT, as Fig. 12 illustrates. The primary frequency component of the sinc function is equal to the modulation frequency. Thus, Millimetro assigns unique OOK modulation frequencies to different tags for simultaneous identification and localization.

However, in practical deployment, the sinc function may be confused by the overlap of the harmonic peaks of the square wave from two different tags. To tackle this challenge, **R-fiducial** [31] designs a novel OOK scheme, called spread spectrum modulation, to guarantee reliable identification, where the distribution of "on" and "off" in the spread spectrum modulation is no longer uniform. In contrast, different OOK sequences are assigned to corresponding tags. The spreading code of each tag is orthogonal to each other, and they can be distinguished by cross-correlating technique.

*6.2.2 FSK.* Although OOK is very simple and widely adopted, however, the amplitude of the received signal at Tx depends on not only the backscatter signal but also reflections from surrounding objects. A proof-of-concept experiment conducted in **UniScatter** [112] demonstrates that if ambient reflections are strong, the Received Signal Strength (RSS) of the backscatter signal will be distorted significantly, leading to the failure of OOK modulation scheme. Therefore, UniScatter chooses FSK as the modulation scheme, which is proven to be more robust against strong ambient reflections. Unlike OOK which uses different amplitude of the backscatter signal to represent "0" or "1", FSK [66, 146, 159] shifts the frequency of the incident signal with two different values to modulate information.

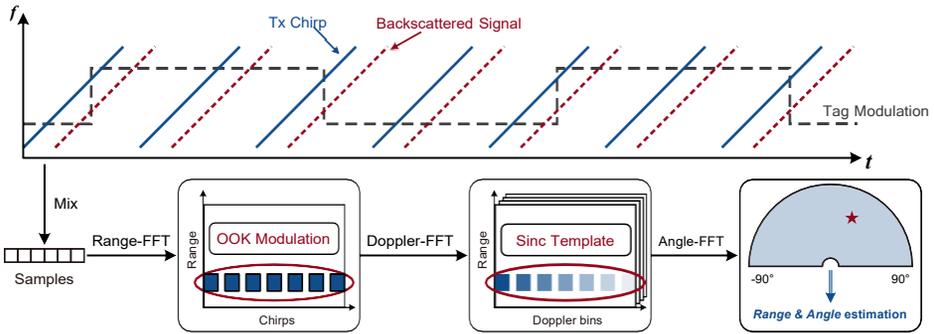

Fig. 12. Concurrent tag localization and identification in Millimetro.

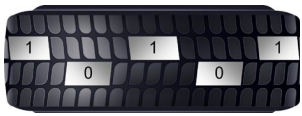

Fig. 13. Spatial coding scheme of Osprey.

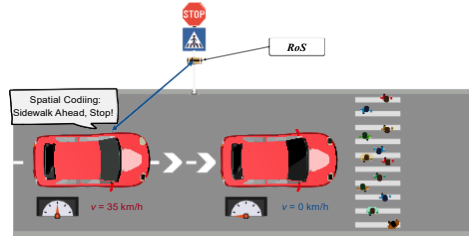

Fig. 14. R2V communication based on spatial coding.

As for UniScatter, the incident signal arriving at the tag will experience a positive/negative frequency shift at a specific switching rate. To decode the backscatter signal, UniScatter fetches all detected peaks and their Doppler frequency spectrum to separate the tag from ambient reflections with matched filtering. Then, given the predefined switching rate of the tag, a Doppler frequency template can be generated. The maximum cyclic cross-correlation value between the template and the Doppler frequency spectrum of each detection can be used as an indicator to detect the tag. Similar to Millimetro, unique modulation frequencies are allocated to tags for concurrent detection.

In order to detect the FSK backscatter with extreme sensitivity, **OmniScatter** [9] proposes High Definition FMCW (HD-FMCW) technique to interplay with tags. Specifically, HD-FMCW leverages multi-chirp symbols to effectively disentangle the ambient reflection from the FSK backscatter signal in the frequency domain. This yields a vast amount of 50dB SNR gain on top of the original FMCW, enhancing the reliability of mmWave backscatter [75, 85, 137].

Further, **Hawkeye** [11] combines HD-FMCW, FSK modulation and planar VAA to achieve hectometer-range subcentimeter localization for large-scale tags. The modulator in Hawkeye's tag is a combination of the hybrid coupler and reflective network that yields a 180° phase flip with path switching. The retro-reflectivity is intact whilst FSK modulation via carefully designed bias and compensation of the coupling effects, avoiding leakage or distortion of the signal.

**BiScatter** [106] can innovatively achieve low-power two-way backscatter communication and sensing, where the uplink modulation is the OOK scheme. As for the downlink communication, the mmWave radar in BiScatter creates multi-bit symbols by varying the FMCW chirp slopes. As such, the downlink signal can be generated by simply changing the radar chirp duration. When this signal goes through the two delay lines with different lengths at the tag, the chirp slope can be estimated according to the baseband beat frequency. Then, the tag can demodulate the downlink data and use either FSK or OOK to modulate the uplink data.

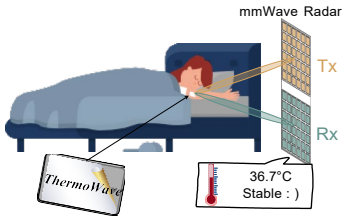 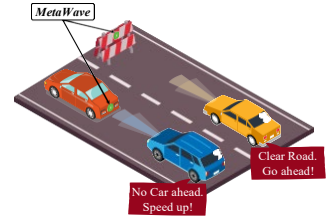

Fig. 15. Operation principle of ThermoWave.  Fig. 16. Typical attack scenarios of MetaWave.

*6.2.3 Spatial Coding.* In many application scenarios, mmWave backscatter transmitters are required to be flexible, environment-friendly, or even pressure-proof. As such, traditional electronic designs are not suitable. In this case, some purely passive designs (*e.g.,* RFID tags) are proposed to meet these requirements. However, as these backscatter transmitters are no longer reconfigurable, their ability is limited. Many existing works utilize them to modulate information based on spatial coding in identification tasks [81, 92, 104, 109]. Specifically, a unique structure is distinctively assigned to each backscatter transmitter to represent its signature, and Tx can identify which one it is querying after receiving the backscatter signal.

**Osprey** [109] proposes a spatial coding scheme to measure tread depth. As Fig. 13 shows, the passive tag is carefully crafted as the layout of the metallic strips in the groove along specific coded patterns. Osprey maps "0" and "1" to different widths of metallic strips. The widths are chosen as 4 cm and 2 cm, which are dictated by the resolution of the mmWave Radar. However, bit errors and collisions between codes from adjacent grooves may happen in this spatial coding scheme. To address this challenge, Osprey borrows from Optical Orthogonal Code (OOC) [29] to maximize the amount of metal and remain within the space constraints of common grooves. It simultaneously achieves two purposes, that is, resilient to collisions and robust to erroneous bits.

**RoS** [104] introduces a spatial encoding-based tag working as an RF barcode to replace road signs. As Fig. 14 shows, tags are constructed by passive VAA stacks, and the information can be modulated in the backscatter signal directly using the layout of tags. An autonomous vehicle can measure the RCS of the tag, from which it estimates the spatial layout and hence decodes the information bits. With such spatial coding, RoS can configure the information bits by altering the number of VAA stacks and adjusting their placement.

**FerroTag** [81] designs a paper-based mmWave-scannable tag for inventory management. FerroTag is a type of passive tag based on FerroRF effects [130, 132]. Specifically, the magnetic nanoparticles within the ferrofluidic ink reply to the incident signal with classifiable features (*i.e.,* the FerroRF response). By designating the ink pattern and hence the location of particles, the related FerroRF response can be modified to form a unique spatial coding. Thus, a specifically designated ferrofluidic ink printed pattern, which is associated with a unique FerroRF response.

*6.2.4 Others.* Besides modulation schemes exploiting explicit rules to modulate the backscatter signal, there are some other types of backscatter transmitters that can piggyback information by creating different signal responses of the backscatter signal.

**ThermoWave** [20] exploits a tag based on metamaterial to passively monitor temperature. Specifically, this tag is composed of a type of metamaterial called cholesteryl materials. The operation principle of this metamaterial is illustrated in Fig. 15. Specifically, when this metamaterial is attached to the target object, it will immediately reach the same temperature as the target object and keep thermal equilibrium according to the theory of thermodynamics [73]. When the temperature of this metamaterial changes, its underlying structure, which contains polymersome (*i.e.,* vessels), will alter its molecular alignment due to thermal expansion [33]. This molecular alignment directly impacts the scattering angle when such metamaterial is probed by broadband mmWave signals. Specifically, when the incident signal arrives at the tag, a frequency shift will be introduced in the backscatter signal, allowing Rx to infer the temperature.

Table 5. Summary of the Performance of Typical Modulation–based mmWave Backscatter Techniques

| Work | Frequency (GHz) | Number of Elements | Energy Efficiency (nJ/bit) | Maximum Range[a] (m) | Data Rate |
|---|---|---|---|---|---|
| **Millimetro** [122] | 24-24.25 | $2 \times 8$ | 2.6-7.9 | 200 | 300-1100 bps |
| **Platypus** [52] | 24-24.25 | $2 \times 8$ | 2.5-4.3 | 100 | 600-1000 bps |
| **mmTag** [99] | 24-24.25 | 6 | 2.4 | 14 | 0.1-1 Gbps |
| **MilBack** [89] | 26.5-29.5 | 1 | 0.5-0.8 | 8 | 36-40 Mbps |
| **BiScatter** [106] | 24-24.25 | $2 \times 8$ | 40-80 | 7 | 50-100 Kbps |
| **OmniScatter** [9] | 24-24.25 | $2 \times 8$ | 583.3 | 40 | 12 bps |
| **UniScatter** [112] | 24-24.25 / 60 ($B$ =3.52)[b] / 77 ($B$ =3.66) | 1 | $2.5 \times 10^6$ | 34 | 20 bps |
| **RoS** [104] | 77-81 | $6 \times 32$ | N/A | 3 | N/A |
| **FerroTag** [81] | 24 | 1 | N/A | 1 | N/A |
| **MilliSign** [54] | 77-81 | $8 \times 8$ | N/A | 15 | N/A |

[a] Tx and Rx are co-located in the architecture of the modulation-based mmWave backscatter.
[b] $B$ stands for bandwidth.

**MetaWave** [25] customizes a series of low-cost tags based on metamaterial, which can be exploited to tamper the modulation of the incident mmWave signal to attack mmWave sensing applications, as depicted in Fig. 16. MetaWave proposes three kinds of tags, specifically: Absorption tag is to attenuate the amplitude of the backscatter signal, which makes the obstacle/intruder ahead disappear. Reflection tag is to rebound the incident signal to create ghost objects and affect the position and speed measurements. Polarization tag is to restrict the specific fields in the incident signal movement to affect the speed and angle measurements.

*6.2.5 Discussion.* We conduct two comparative analyses, performance comparison and modulation scheme comparison, to gain insights for modulation-based mmWave backscatter.

**Comparison between typical modulation-based works.** Table 5 summarizes the key performance of typical modulation-based techniques, including the frequency, number of elements in the backscatter transmitter, energy efficiency, maximum range and data rate.

Most modulation-based works exploit the backscatter transmitters operating below 30 GHz, due to the design challenges of the RF switch in high frequency. Alternatively, passive backscatter transmitters can be employed to achieve spatial coding in high frequency, such as **RoS** and **MilliSign**. In contrast, **UniScatter** achieves FSK in 60 GHz and 77 GHz bands. The main reason is that the LL-based tag designed in UniScatter is made of graphene material, which shows wideband reconfigurability [112].

However, the cost of UniScatter is a low energy efficiency of $2.5 \times 10^6$ nJ/bit due to the very large capacitance of the material. This problem can be tackled by using more low-power materials and reducing the size of the tag. **OmniScatter** also exhibits a low energy efficiency. This is because multiple FMCW frames are combined in a symbol to enhance the frequency resolution and sensitivity, thus degrading the data rate and the energy efficiency. For most other works, the energy efficiency is under 100 nJ/bit.

One of the most significant enhancements brought by mmWave backscatter is the operation range. Many works increase the maximum range to above 10 m by fully exploiting the retro-reflective backscatter tags to enhance the energy of backscatter signals. **Millimetro** [122] and **Platypus** [52] can even work at over 100 m by detecting the unique features in the backscatter signal, but at the cost of relatively low data rate. In comparison, **FerroTag** [81] shows a shorter operation range due to the tags do not retro-reflect the signal and require a considerable SNR to ensure the identification accuracy.

Besides, a shorter operation range generally contributes to a higher data rate, but it is also impacted by other factors, such as the modulation scheme. Also, the data rate may be sometimes intentionally controlled to lower the total energy consumption.

**Comparison between typical modulation schemes.** Different modulation techniques have distinct advantages and limitations, which suit the requirements of different applicable scenarios.

- **OOK**: The main advantage of OOK modulation is its simplicity. This makes it easy to implement and understand. Besides, OOK modulation requires minimal power for transmission and does not need complex hardware. However, the very simplicity of OOK is also a limitation. It is highly susceptible to noise and interference. Ambient reflections or other environmental variations can cause heavy signal distortion, leading to data interpretation errors. Furthermore, its data rate may be also limited. Thus, OOK modulation is suitable for systems where complexity needs to be minimized, such as resource-constrained devices.
- **FSK**: FSK modulation's main advantage lies in its robustness against amplitude noise and interference. Thus, it can provide more robustness and stable performance in mmWave backscatter systems. However, FSK requires more sophisticated hardware and signal processing, which can lead to increased energy and hardware costs, especially at such a high mmWave band. As such, FSK-based mmWave backscatter is more suitable for distinguishing data transmission from different backscatter transmitters, where device and energy costs are not the main concerns.
- **Spatial Coding**: Spatial coding benefits from quick and low-cost fabrication because it does not need electronic modules. However, the drawback of this type of modulation scheme is the limited reconfigurability. Once a tag is physically constructed, it is hard to alter its design or the encoded information. Therefore, spatial coding is most applicable in identification applications, where identifiers do not need frequent updating.

## 6.3 Signal Relay

When sources of energy and information are both Tx, backscatter transmitters will mainly work as a relay to redirect the incident signal to another Rx, and this is the architecture of relay-based mmWave backscatter. Table 6 summarizes relay-based works, among which there are typically two relay schemes according to whether the radiation pattern of backscatter transmitters can be changed, that is, reconfigurable pattern and constant pattern. The following will introduce these relay techniques and corresponding works, respectively.

*6.3.1 Reconfigurable Pattern.* In the context of relay-based mmWave backscatter, the precise alignment of the backscatter signal becomes paramount, particularly when considering the mobility of Rx. Given that Rx terminals are often mobile devices, and in some scenarios, there may be multiple Rx terminals to communicate with, it necessitates dynamic or multiple beams to ensure effective communication. To address this challenge, backscatter transmitters, in this case, are expected to be reconfigurable to dynamically adapt the beam of the backscatter signal.

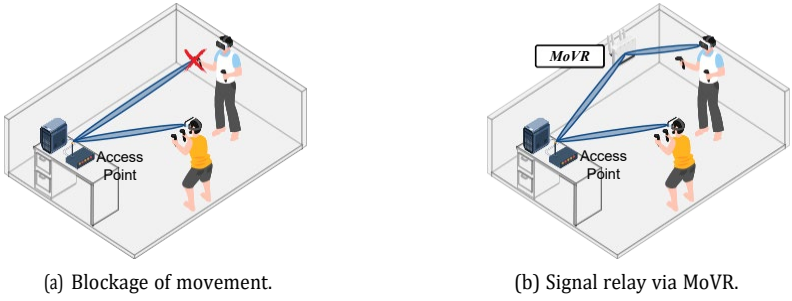

(a) Blockage of movement.  (b) Signal relay via MoVR.

Fig. 17. Application scenarios of MoVR.

Table 6. Summary of the Works of Relay–based mmWave Backscatter

| Works | Relay Scheme | Backscatter Transmitter Design | Application Scenario |
|---|---|---|---|
| **MilliMirror** [111] | Constant Pattern | Metasurface | Outdoor coverage expansion |
| **GigSky** [76] | Constant Pattern | Leaky Wave Antenna | Outdoor coverage expansion |
| **AutoMS** [94] | Constant Pattern | Metasurface | Indoor coverage expansion |
| **Zhang *et al.*** [148] | Reconfigurable Pattern | Metasurface | Indoor coverage expansion |
| **MoVR** [1] | Reconfigurable Pattern | Phased Array | Indoor coverage expansion |
| **mmXtend** [77] | Reconfigurable Pattern | Leaky Wave Antenna | Outdoor coverage expansion |
| **mmWall** [28] | Reconfigurable Pattern | Metasurface | Indoor coverage expansion |
| **MeSS** [22] | Reconfigurable Pattern | Metasurface | Attacking |

**MoVR** [1] focuses on using mmWave backscatter to deliver multi-Gbps wireless communication between VR headsets and their game consoles. Since mmWave radios use highly directional antennas, they work only when the Tx's (or tag's) beam is aligned with the Rx's beam. Therefore, as Fig. 17 illustrates, even a small movement of the headset can hamper the alignment and break the link. To adapt to this mobility, MoVR introduces a self-configurable phased array-based tag that detects the incident signal and reconfigures itself to backscatter it toward the Rx on the headset. In contrast to the traditional tag, a MoVR tag does not require the angle of backscatter to be equal to the angle of incidence. Both angles can be programmed so that it can receive the signal from the mmWave Tx attached to the data source (*i.e.,* the game console) and backscatter it towards the player's headset, regardless of its direction. This kind of tag can be implemented simply by deflecting the analog signal without any decoding.

**mmWall** [28] proposes a metasurface to bolster the reliability of mmWave communication indoors and outdoors. mmWall is designed to be electronically reconfigurable to either reflect or refract incoming energy, allowing it can either reflect the beam of the incident signal at non-specular to bypass the indoor blockage or backscatter the outdoor signals at non-specular angles to support multiple indoor users at the same time. mmWall can time-multiplex each of the above use cases without human intervention, while installed in a fixed location. Moreover, mmWall has no RF chain, and its electric components draw only a couple of hundred $\mu$W orders of power.

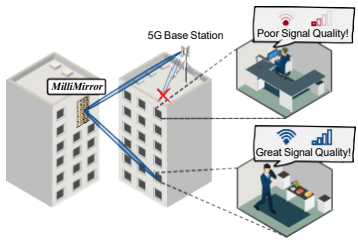
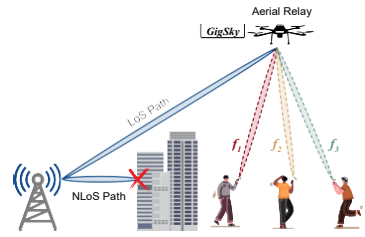

Fig. 18. Typical use case of MilliMirror.           Fig. 19. Example scenario of GigSky.

*6.3.2 Constant Pattern.* To achieve reconfigurable patterns, the hardware implementation is usually complex and bulky, not adequate for some scenarios that require lightweight and low-cost solutions. As such, backscatter transmitters with constant patterns are proposed. They are often convenient to be deployed indoors and outdoors, and even with the aid of other mobile devices to support dynamic Rx. Moreover, many of these backscatter transmitters with constant patterns may be quickly fabricated with the development of PCB and 3D printing technology.

Due to the blockage of LoS and lack of (or weak) multipath from natural reflectors, signals from the outdoor mmWave base station can not reach the indoor user equipment. To tackle this problem, **MilliMirror** proposes a 3D printed metasurface, which is totally passive, to achieve the constant relay pattern. Specifically, the unit of MilliMirror metasurface is a type of metal-backed dielectric cuboid (MBDC). By tuning the thickness of the unit, different phase shifts ranging from 0° to 360° can be applied to the backscattered signal. As the MilliMirror metasurface is composed of thousands of MBDC units, the metasurface can delicately fine-tune the thickness and distribution of the units to redirect the mmWave signal to any direction. Moreover, as the structure of the MilliMirror metasurface is passive and can not be changed once it is fabricated, it can provide a constant relay pattern after deployment. The typical use case of MilliMirror is demonstrated in Fig. 18. By placing the MilliMirror metasurface nearby to relay the mmWave beam, a strong NLoS path can be created to illuminate the coverage blind spots. Unlike active relays or digitally controlled reflectarrays, it has no electronic components or integrated circuits (ICs), and can be mass fabricated through low-cost 3D printing processes.

In comparison, **AutoMS** [94] proposes an automated service framework to optimize indoor mmWave coverage with networked passive metasurfaces. A joint optimization method is exploited to decide the optimal setting of the metasurface's phase configuration and placement as well as the AP's beamforming codebooks and placement. As a result, the AP's dynamic beam steering capability and the passive metasurface's phase-shifting capability are combined together to enhance the room-scale signal coverage and received signal strength.

**GigSky** proposes a LWA-based tag that utilizes unmanned aerial vehicles (UAVs) as aerial relays to provide over-the-air non-blockage communication paths between Tx and Rx, as Fig. 19 shows. To cover all users on the ground and simultaneously provide connectivity with high date rate to them, GigSky creates multiple beams where each points to a user. Further, to minimize the interference between adjacent users, each beam should operate on a different frequency channel. GigSky achieves this by carrying a LWA on the UAV.

*6.3.3 Discussion.* We conduct two comparative analyses, performance comparison and relay scheme comparison, to gain insights for relay-based mmWave backscatter.
**Comparison between typical relay-based works.** Table 7 summarizes the key parameters and performance of typical relay-based techniques, including the number of units in backscatter transmitters, coverage and performance enhancement brought by mmWave backscatter.

Table 7. Summary of the Performance of Typical Relay–based mmWave Backscatter Techniques

| Works | Num. of Units | Coverage (m × m) | Performance Enhancement |
|---|---|---|---|
| **MilliMirror** [111] | 80 × 80 | ~3 × 3 | 5 dB gain |
| **AutoMS** [94] | 160 × 200 | 13.8 × 9.2 | 11 dB gain in target rooms 20 dB gain at blind spots |
| **Zhang** *et al.* [148] | 16 × 16 | 6 × 8 | 10 % coverage increase |
| **MoVR** [1] | 8 | ~3 × 3 | 14 dB gain |
| **mmXtend** [77] | 8 × 17 | ~10 × 8 | 20 dB gain |
| **mmWall** [28] | 28 × 76 | ~7 × 8.75 | 30 dB gain |

The backscatter transmitters in relay-based mmWave backscatter usually rely on large number of units to change the propagation direction of the incident signals (*e.g.,* beam steering) to relay the signal from the Tx to the Rx. For instance, the passive metasurface proposed in **MilliMirror** [111] and **AutoMS** [94] are composed of thousands of units to achieve constant relay patterns. However, by using the reconfigurable metasurface, **mmWall** [28] can achieve more dynamic and fine-grained manipulation of the signal, so it can reduce the number of units to achieve a comparable performance. Similarly, other backscatter transmitters with reconfigurable patterns (*e.g.,* phased array in **MoVR** [1]) also show a limited number of units.

With these backscatter transmitters, the coverage of mmWave signal can be significantly enhanced compared with the traditional mmWave communication. Traditional mmWave communication usually struggles to reach blind spots due to indoor/outdoor blockages. In contrast, as Table 7 shows, relay-based mmWave backscatter technology can extend signal coverage to tens, or even over a hundred square meters, achieving signal gains of tens of dB, thereby enabling room-scale coverage enhancement.

**Comparison between typical relay schemes.** Different relay techniques have respectively distinct advantages and limitations, which suit the requirements of different applicable scenarios.

- **Reconfigurable pattern**: The reconfigurability has the distinct advantage of dynamically adjusting beam directions or other signal attributes. This feature allows mmWave backscatter transmitters to maintain optimal alignment with moving Rx devices. Additionally, the reconfigurability shows the potential to serve multiple users, thereby enhancing versatility and ensuring robust performance in complex, multi-user environments. However, reconfigurable patterns require sophisticated hardware and control mechanisms, which increase design and operational complexity. Consequently, they are generally more costly and show higher power consumption. As such, mmWave backscatter systems with reconfigurable patterns are more suitable for infrastructures to support environments with high mobility and frequent positional changes, such as urban settings or vehicular networks.
- **Constant pattern**: Constant patterns offer a simple and cost-effective way of designing and deploying mmWave backscatter transmitters. Specifically, advancements in fabrication technologies, such as PCBs and 3D printing, empower rapid manufacturing of constant pattern backscatter transmitters, facilitating quick deployment and scalability. The ease of fabrication and deployment, however, comes with limitations. Constant patterns lack flexibility and are inherently static, which restricts their adaptability in dynamic environments. In this case, effective communication and sensing may be not feasible. When considering the optimal application scenarios, constant patterns are suited for static or predictable environments, such as warehouses or fixed industrial settings, where the relative positions of Tx and Rx remain stable.

## 7 CHALLENGES AND POTENTIAL DIRECTIONS

Previous discussion has highlighted the recent progress and emerging opportunities under the topic of mmWave backscatter. However, there still exist several challenges in this area. In this section, we will introduce a few remaining challenges and the corresponding promising directions to deal with them. We believe that these directions will be investigated in future research to boost the development of mmWave backscatter.

### 7.1 Optimizing Hardware Platforms

*7.1.1 Challenge.* Although existing efforts make some breakthroughs in mmWave backscatter, the current technologies and applicability still partly suffer from limited ability of mmWave hardware platforms. On the one hand, most platforms are a little bit bulky and tied by connected cables. Taking one of the most used FMCW radar, TI IWR6843 [60], as an example, there are in total four cables needed, in which two for power supply and two for communication with the computer. This cumbersome deployment hinders the convenience and scalability of mmWave backscatter, particularly in mobile scenes. On the other hand, COTS platforms are favored among lots of works, but they can not well match the requirements of some specific backscatter applications. For instance, RoS [104] exploits polarization switching to cancel the interference between the incident and backscatter signal. However, the COTS radars usually only equip with linear polarized antennas. As such, two radars are needed to put RoS prototypes into practical use.

Based on the above observation, we envision two promising directions to optimize mmWave hardware platforms. We hope to provide researchers with some ideas and reasonable imagination to promote innovation in mmWave platforms and backscatter techniques.

*7.1.2 Direction 1: Miniaturized Platforms for Mobile Scenes.* Extensive attempts have been made to miniaturize mmWave platforms [84, 105, 108, 145] aided by either complementary metal-oxide-semiconductor (CMOS) or radio-frequency integrated circuit (RFIC) technique. As such, the RF front-end and some other components of mmWave platforms can be packaged into a chip with a very small size. We can envision that, these miniaturized platforms are able to support mobile backscatter applications conveniently, and further, they can be even integrated with our daily mobile devices, such as phones, pads, laptops and so on to boost the pervasive mmWave backscatter. Note that there is indeed a mature product, called **Soli** [84], which works at 60 GHz ISM band. The Soli chip incorporates the entire sensor and antenna array into an ultra-compact 8 mm $\times$ 10 mm package. This miniaturization helps Soli to be shipped in a phone and other mobile devices.

Even though there are not other consumer devices proposed after Soli, we witness continuous attempts of researchers and engineers to step toward to vision of mmWave platform miniaturization. We believe that this will be a major trend to make pervasive mmWave backscatter come true.

*7.1.3 Direction 2: Customized devices for Different Demands.* Flexible customization could inspire more innovative technologies and applications in mmWave backscatter. Besides existing COTS mmWave platforms that only support fixed functions without customization in the hardware layer, we have already seen many customized devices put into use (ref to Sec. 5). This provides many benefits that can satisfy unique requirements in some application scenarios. Although some works can use the signal generator and spectrum analyzer with a specific antenna to provide much convenience, the economic cost may be much higher. Flexible customization not only stimulates the community creatively, but also provides much choice for researchers. We believe the efforts in this direction will largely inspire the area of mmWave backscatter.

**7.2 Innovating Backscatter Transmitters**

*7.2.1 Challenge.* Backscatter transmitters (tags/metasurfaces) are the core of mmWave backscatter technology. In many works, the most challenging task and the most valuable contribution lies in the backscatter transmitter design. However, some backscatter transmitters, even fancy and useful, are hard to put into practical deployment. The reason is two-fold, complex fabrication and high energy consumption. From the perspective of fabrication, some backscatter transmitters are composed of various components and require very careful and tangled cable connections, *e.g.,* metasurface. This leads to complex debugging and simulation before real manufacturing, because hundreds of parameters, components and materials are involved. Even after the successful theoretical design, a long time is required for the production process. And finally, it must be careful to assemble and debug the backscatter transmitter in practice, which also may be time-consuming and error-prone. From the perspective of energy consumption, some of the existing backscatter transmitters are also a little bit power-hungry but do not incorporate any energy harvesting schemes. In that case, these backscatter transmitters and the related techniques are far away from practical deployment.

To achieve pervasive mmWave backscatter in IoT or 5G networks, we think there are following three promising directions of future research: quick fabrication, power reduction and integrating energy harvesting modules with backscatter transmitters.

*7.2.2 Direction 1: Quick Fabrication via 3D Printing Process.* Considering the complexity of current mmWave backscatter transmitters, various methods are proposed to simplify the fabrication process [25, 81, 111, 119, 139]. Among them, 3D printing [14–17, 111] may be one of the most promising technologies. Recently, 3D printing has extended from conventional polymer to general materials, including various dielectric and metal. Thus, exploiting 3D printing to fabricate RF components has been as an economic and convenient replacement of PCB. People can only wait for several hours after they feed the design of devices into 3D printer, such as lenses [15], reflectarrays [21], waveguides [14] and so on. Moreover, we have already witnessed the 3D printed mmWave metasurface proposed in MilliMirror [111]. Even though 3D printed metasurface benefits from quick fabrication and low cost, its advantage is the non-reconfigurable structure. Once they are produced, their radiation patterns are hard to change. Thus, we envision that prefabricated components will be the attractive direction. What kind of backscatter transmitters do you need to design, you can use what kind of prefabricated components to assemble. In that case, we can reconfigure backscatter transmitters like building blocks.

*7.2.3 Direction 2: Power Reduction using Passive Elements.* Traditional mmWave backscatter transmitters, like phased arrays, consume much energy. For example, MoVR [1] builds a backscatter transmitter based on the phased array called MiRa [2]. The power consumption of MiRa is 11.6W [99], which is so high for the resource-limited backscatter transmitter. This is caused by the power-hungry RF chain used in the backscatter transmitter. Recent progress in mmWave backscatter transmitters tries to use analog elements to replace these RF chains. For example, mmTag [99] exploits a passive retro-reflective tag to achieve the beamforming function of phased arrays, reducing four-fifths of energy. Further, mmWall [28] reforms the design of metasurface in the mmWave band. It only uses passive antennas to reflect or absorb signals and does not need RF chains for beam control. Thus, its electric components draw only a couple of hundred microwatts order of power. With the development of mmWave backscatter, the mainstream may be low-power digital modules for control and passive elements for signal manipulation. As such, the power of mmWave backscatter transmitters can be reduced to achieve practical batteryless backscatter.

*7.2.4 Direction 3: Integrating Energy Harvesting Modules.* To support the continuous operation of mmWave backscatter transmitters, it is essential to incorporate energy harvesting modules,

# A Survey of mmWave Backscatter: Applications, Platforms and Technologies 27

particularly for modulation-based mmWave backscatter systems. These modules provide the necessary power for both startup and ongoing internal operations. However, many existing mmWave backscatter systems have not integrated energy harvesting technologies into their backscatter transmitter design yet. Drawing inspiration from sub-6 GHz backscatter systems, which commonly harvest energy from incident signals, future mmWave backscatter could adopt a similar strategy.

As the mmWave signal is generally transmitted in a beamforming pattern, the above-mentioned strategy can achieve higher energy efficiency than that in sub-6 GHz systems [45, 110, 131]. However, the high path loss of mmWave signals imposes a limitation on the effective range of energy harvesting. Additionally, the issue of beam alignment also concerns the efficiency of this strategy. To avoid these challenges, an alternative approach is to integrate supplementary sensors for energy harvesting, such as polar, kinetic and piezoelectric sensors [93], which could enhance the robustness and versatility of energy harvesting solutions in mmWave backscatter systems.

## 7.3 Exploring Practical Deployment

*7.3.1 Challenge.* Researchers and developers in the area of mmWave backscatter hope to put this technology into practical use and deployment. However, the current prototypes are far from the real deployment and face many practical challenges. First, from the perspective of cost-effectiveness, mmWave backscatter needs dense deployment of both mmWave platforms (*e.g.,* mmWave radars or APs) and backscatter transmitters, which may require large expenditure to establish the infrastructure. Second, from the perspective of applications, most works either achieve mmWave backscatter-based communication or sensing, but are not yet able to combine these two tasks to achieve ISAC well. Besides, existing mmWave backscatter-based applications are ad hoc with different platforms and frequency bands. It is hard to standardize or integrate these applications into a backscatter network or communication standards. Third, from the perspective of techniques, many important designs for a backscatter system are often overlooked by existing prototypes, such as interference mitigation in backscatter systems as well as signal detection and channel estimation. Based on the above discussion of the challenges, we elaborate on the possible directions for practical deployment from three aspects, namely cost-effectiveness, applications (*i.e.,* ISAC, backscatter networks, and ground-air cooperation) and techniques (*i.e.,* interference mitigation, signal detection and channel estimation, and standard compatibility)

*7.3.2 Direction 1: Enhancing Cost-Effectiveness of Deployment.* The cost of mmWave platforms and backscatter transmitters is significantly higher than that of conventional sub-6 GHz backscatter systems. Additionally, the limited coverage range of mmWave backscatter necessitates a densely packed deployment to achieve large-area coverage, which further escalates costs. To enhance the cost-effectiveness of mmWave backscatter in practical deployments, efforts could be two-fold. Firstly, we can focus on specific application scenarios that benefit from high frequency and throughput, such as VR/AR so that reducing the need for extensive, large-scale deployment. Secondly, we can leverage new materials and manufacturing technologies to help simplify device design and lower the production costs of mmWave backscatter devices. Moreover, standardizing mmWave backscatter systems and integrating them with existing infrastructures can maximize cost-effectiveness by avoiding redundant investments and enabling seamless operation within established networks.

*7.3.3 Direction 2: Integrated Sensing and Communication.* ISAC is promising for future IoT technologies [86, 91, 95], while some of the mmWave backscatter techniques have already shown the potential for ISAC. For example, mmTag [99] and RoS [104] are designed based on VAA, which can be accurately localized because of the high energy of the backscatter signal. Thus, it is possible to integrate localization ability in these two works. After that, those methods can be further extended to be combined with the state-of-the-art micro-displacement estimation technique [52] for mobile

, Vol. 1, No. 1, Article . Publication date: April 2024.

tag tracking. This may increase power consumption by requiring a higher switching rate, but offers substantial promise for future ISAC-enabled IoT networks.

Millimetro [122] and Hawkeye [11] combine the communication and identification abilities in the design of tags, which seems to achieve ISAC. However, tags in these two works can only modulate the incident signal with a fixed OOK frequency. In other words, the communication process can not exchange any information except the ID of the tag. By exploiting more signal properties, such as RCS, phase variation and polarization, they may gain more communication capability and provide more information with enhanced sensing performance.

*7.3.4  Direction 3: Backscatter Networks and Systems.* Although current research in mmWave backscatter has contributed many innovative techniques, valuable theoretical frameworks and proof-of-concept demonstrations, they are still scattered. This scattered approach has its limitations that often focus on specific scenarios, but may not address the broader challenges of deploying mmWave backscatter in the real world. To bridge this gap, it is imperative to establish dedicated mmWave backscatter networks and systems. In these networks and systems, different applications, platforms and technologies are put together, among which problems or opportunities may be inspired, so research in this direction is very promising for mmWave backscatter.

*7.3.5  Direction 4: Ground-Air Cooperation.* One of the key challenges in mmWave backscatter systems is maintaining uninterrupted communication paths in varied and obstruction-rich settings. The inherent characteristics of mmWave frequencies, such as limited penetration ability and susceptibility to blockages, demand innovative solutions to ensure consistent and reliable performance. Ground-Air cooperation emerges as a promising approach [76, 124]. By employing UAVs equipped with advanced mmWave relay capabilities, networks can dynamically navigate and avoid obstacles, thus maintaining robust LoS paths. This mobility allows aerial platforms to act as dynamic relays, continually optimizing their positions to sustain high-quality linkages with ground-based users and devices. Furthermore, integrating ground-air cooperative mmWave backscatter with existing communication infrastructure holds substantial potential for enhancing network capabilities. For instance, this hybrid setup can facilitate load balancing across the network by dynamically rerouting traffic through aerial relays. There is ongoing research aimed at exploring this area [76]. We believe that continuous exploration in this direction can lead to significant advancements in communication infrastructures and mmWave backscatter systems.

*7.3.6  Direction 5: Interference Mitigation in Backscatter Systems.* Backscatter systems may suffer from the interference of the reflection from ambient facilities. Existing efforts to deal with this interference are exploiting polarization switching [99, 104]. Indeed, it has been approved to be an effective way. However, this method requires two mmWave devices, one for transmitting and one for receiving, because almost all existing mmWave devices are linearly polarized. Besides, there may be a half energy loss of the backscatter signal due to the polarization mismatch, hindering the detection and receiving of the backscatter signal. The promising way to address this issue may be exploring circular polarized or configurable antennas for updating mmWave devices.

Besides, mmWave backscatter networks often involve multiple devices that transmit or reflect signals simultaneously. This can introduce interference between devices, significantly degrading system performance. While some existing studies have assigned unique modulation patterns for different tags to distinguish between them [11, 122], these approaches do not fully address situations where multiple mmWave Txs are involved, leading to potential performance bottlenecks. To mitigate interference in these networks, future mmWave backscatter systems can incorporate Time Division Multiple Access (TDMA) or Carrier Sense Multiple Access (CSMA) schemes, similar to those used in conventional sub-6 GHz backscatter networks [158, 160].

*7.3.7 Direction 6: Signal Detection and Channel Estimation.* Most of the existing mmWave backscatter systems lack practical schemes for signal detection and channel estimation, which are crucial for improving the quality of backscatter-based communication and sensing.

On one hand, accurately detecting and decoding the incident signals could help downlink communication. However, the limited computation and energy resources of backscatter transmitters present challenges in achieving this goal. Thus, most existing approaches only facilitate uplink communication and often overlook the downlink communication. Recent works, **MilBack** [89] and **BiScatter** [106], take the first step to deal with this challenge. MilBack exploits an FMCW radar to send a "V-shape" signal, and the LWA-based tag will preserve two narrowband signals with the same frequency based on its frequency selectivity. An envelope detector is then used for signal detection and estimates the delay of these two signals for frequency identification. BiScatter employs an FMCW radar to send signals with different slops, and the tag employing two delay lines of different lengths can estimate the slope for signal detection and decoding. Although both approaches require dedicated waveforms of the incident signals, they provide an important insight that we can leverage the analog features of various mmWave backscatter tags to replace the complex digital signal processing. This insight opens up new possibilities for signal detection and decoding.
On the other hand, channel estimation is also important for optimizing communication links. The unique characteristics of mmWave frequencies, such as high path loss and sensitivity to obstructions, necessitate precise channel estimation to enhance signal quality and system performance. However, this topic remains largely unexplored in the existing mmWave backscatter systems. A promising direction is to leverage existing standards, such as IEEE 802.11ay and IEEE 802.11ad, which provide related techniques for channel estimation (*e.g.,* channel estimation field in IEEE 802.11ad [19]) that may be adapted for mmWave backscatter applications. Future exploration could first try to use 802.11ad/ay-compliant platforms for channel estimation at Rx side, and then extend to the backscatter transmitter side.

*7.3.8 Direction 7: Compatibility with Existing Communication Standards.* To push mmWave backscatter technology from prototype to deployment, one of the promising steps is to keep compatible with existing communication standards (*e.g.,* IEEE 802.11ad/ay). However, most of the existing mmWave backscatter systems are implemented with FMCW radar or customized devices, which limits their integration potential with mainstream networks. Future exploration can take inspiration from sub-6 GHz backscatter systems, which effectively embed backscatter information by leveraging the distinctive features of communication standards. For example, in mmWave band, one may use beamforming training frames that are frequently exchanged between 802.11ad/ay APs and clients [19] for backscatter information embedding. This can facilitate the seamless integration of mmWave backscatter technology into current network infrastructures.

## 8 CONCLUSION

This paper provides a comprehensive survey on the recent progress in mmWave backscatter technology. We propose a novel taxonomy to categorize existing works into modulation-based and relay-based ones, based on our insight in the different operating principles of backscatter transmitters. Along with a detailed discussion of existing works, emerging applications, hardware platforms, and key technologies are introduced systematically. Afterward, we summarize the quantified the performance of typical works within each category, and analyze the performance enhancements brought by mmWave backscatter techniques compared to traditional mmWave technologies. Moreover, we analyze the advantages, limitations, and applicable scenarios for both modulation and relay schemes. Furthermore, we explore 12 potential future research directions in mmWave backscatter from three perspectives, optimizing hardware platforms, innovating backscatter transmitters, and exploring practical deployment. As a promising technology for high-speed communication and fine-grained sensing, mmWave backscatter is deemed an extension of the current IoT to be further explored in the coming years. We hope this survey will inspire researchers and developers to achieve further progress in the field of mmWave backscatter.

## 9 ACKNOWLEDGMENT

This work is supported by the National Natural Science Foundation of China under grant No. 62425207 and No. U21B2007.